\documentclass[12pt,preprint]{aastex}
\shorttitle{Observations of Arp 299}
\shortauthors{Sliwa $\&$ Wilson}

\begin{document}
\title{Luminous Infrared Galaxies With the Submillimeter Array. \\ III. The Dense Kiloparsec Molecular Concentrations of Arp 299}
\author{Kazimierz Sliwa\altaffilmark{1},  Christine D. Wilson\altaffilmark{1}, Glen R. Petitpas\altaffilmark{2}, Lee Armus\altaffilmark{3}, Mika
Juvela\altaffilmark{4}, Satoki Matsushita\altaffilmark{5}, Alison B. Peck\altaffilmark{6}, Min S. Yun\altaffilmark{7}  }
\altaffiltext{1}{Department of Physics and Astronomy, McMaster University, Hamilton, ON L8S 4M1 Canada; sliwak@mcmaster.ca, wilson@physics.mcmaster.ca}
\altaffiltext{2}{Harvard-Smithsonian Center for Astrophysics, Cambridge, MA 02138; gpetitpa@cfa.harvard.edu}
\altaffiltext{3}{Spitzer Science Center, California Institute of Technology, Pasadena, CA 91125; lee@ipac.caltech.edu}
\altaffiltext{4}{University of Helsinki Observatory, Finland; mika.juvela@helsinki.fi}
\altaffiltext{5}{Academia Sinica Institute of Astronomy and Astrophysics, Taipei 106, Taiwan; satoki@asiaa.sinica.edu.tw} 
\altaffiltext{6}{Joint ALMA Office, Avda El Golf 40, piso 18, Santiago, Chile 7550108; apeck@alma.cl}
\altaffiltext{7}{Department of Astronomy, University of Massachusetts, Amherst, MA 01003;myun@astro.umass.edu}

\begin{abstract}
We have used high resolution ($\sim$2.3") observations of the local ($D_{\rm{L}}$ = 46 Mpc) luminous infrared galaxy Arp 299 to map out the physical properties of the molecular gas which provides the fuel for its extreme star formation activity. The $^{12}$CO J=3-2, $^{12}$CO J=2-1 and $^{13}$CO J=2-1 lines were observed with the Submillimeter Array and the short spacings of the $^{12}$CO J=2-1 and J=3-2 observations have been recovered using James Clerk Maxwell Telescope single dish observations. We use the radiative transfer code RADEX to estimate the physical properties (density, column density and temperature) of the different regions in this system. The RADEX solutions of the two galaxy nuclei, IC 694 and NGC 3690, are consistent with a wide range of gas components, from warm moderately dense gas with $T_{\rm{kin}}$ $>$ 30 K and $n$(H$_{\rm{2}}$) $\sim$ 0.3 - 3 $\times$ 10$^{3}$ cm$^{-3}$ to cold dense gas with $T_{\rm{kin}}$ $\sim$ 10-30 K and $n$(H$_{2}$) $>$ 3 $\times$ 10$^{3}$ cm$^{-3}$. The overlap region is shown to have a better constrained solution with $T_{\rm{kin}}$ $\sim$ 10-50 K and $n$(H$_{2}$) $\sim$ 1-30 $\times$ 10$^{3}$ cm$^{-3}$. We estimate the gas masses and star formation rates of each region in order to derive molecular gas depletion times. The depletion times of all regions (20-60 Myr) are found to be about 2 orders of magnitude lower than those of normal spiral galaxies. This rapid depletion time can probably be explained by a high fraction of dense gas on kiloparsec scales in Arp 299. We estimate the CO-to-H$_{\rm{2}}$ factor, $\alpha_{\rm{co}}$ to be 0.4 $\pm$ 0.3 (3 $\times$ 10$^{-4}$/ $x_{\rm{CO}}$) $M_{\odot}$ (K km s$^{-1}$ pc$^{2}$)$^{-1}$ for the overlap region. This value agrees well with values determined previously for more advanced merger systems.
\end{abstract}

\keywords{galaxy:  individual(Arp 299, NGC 3690, IC 694), infrared: galaxy}

\section{Introduction}
Ultra/Luminous infrared galaxies (U/LIRGs) are systems in the local universe that exhibit extreme star formation. They emit a large portion of their total luminosity at far-infrared (FIR) wavelengths (LIRGs: $L_{\rm{FIR}}$ $\sim$ 10$^{11-12}$ $L_{\odot}$; ULIRGs: $L_{\rm{FIR}}$ $>$ 10$^{12}$ $L_{\odot}$; Sanders $\&$ Mirabel 1996)\nocite{1996ARA&A..34..749S}. Early optical studies have shown that a large fraction of U/LIRGs show morphologies that resemble systems that are interacting or merging \citep{1987AJ.....94..831A,1988ApJ...325...74S}. High resolution optical and near-infrared imaging has revealed that a large fraction of U/LIRGs have nuclear separations of 0.3 - 48 kpc \citep{1996AJ....111.1025M}. The merger process likely triggers the extreme star formation in these galaxies (Sanders et al. 1988) but the detailed process is still not well understood. 
 
U/LIRGs have been observed to contain a large amount of molecular gas with $M$(H$_{2}$) up to about 10$^{10}$ $M_{\odot}$ \citep{1986ApJ...305L..45S}. The molecular gas is seen to be concentrated near the nuclear regions within a radius of 0.5 kpc \citep{1996ARA&A..34..749S}.  Subarcsecond resolution CO observations have revealed rotating disks of molecular gas that has been driven to the nuclei \citep{1998ApJ...507..615D}. Theoretical models have shown that during the collision, the gas loses angular momentum due to dynamical friction which causes the gas to decouple from the stars and flow inwards towards the nuclei \citep{1991ApJ...370L..65B}. The concentrated molecular gas provides fuel for star formation and/or for an active galactic nucleus (AGN). 

\citet{2009ApJ...699..667M} have shown that submillimeter galaxies (SMGs) have similar radiation environments to local starbursts. The distribution of the optical depth determined from the silicate absorption feature at 9.7 $\mu$m was found to be lower than for local U/LIRGs, which suggests that the mid-infrared regions of SMGs have a lower dust obscuration. The difference in extinction between U/LIRGs and SMGs likely arises from differences in geometry. Using $^{12}$CO J=3-2 observations, Iono et al. (2009; Paper II)\nocite{2009ApJ...695.1537I} have shown that the molecular gas disks in local U/LIRGS are more compact (0.3 - 3.1 kpc) than the SMGs (3 - 16 kpc). \citet{2009ApJ...699..667M} also found that the majority of SMGs are dominated by intense star formation just like local U/LIRGs. The differences suggest that ULIRGs are not exact local analogs of high redshift SMGs; however, there are enough similarities that local ULIRGs are often used as templates for SMGs. 

Arp 299 (NGC 3690 +  IC 694, Mrk 171, VV 118, IRAS 11257+5850) is one of the nearest ($D_{\rm{L}}$ = 46 Mpc) examples of a merger system. It has a far-infrared luminosity of $L_{\rm{FIR}}$ = 5.5 $\times$ 10$^{11}$ $L_{\odot}$ (Sanders et al., 2003)\nocite{2003AJ....126.1607S}, classifying it as a LIRG. In the optical, the two nuclei are still distinguishable. High resolution observations show the nuclei are separated by 22$\arcsec$ \citep[4.9 kpc ;][]{1991ApJ...366L...1S}. Gehrz et al. (1983)\nocite{1983ApJ...267..551G} showed that there is also a third region of emission thought to be associated with the overlap of the two galaxies. \citet{1999AJ....118..162H} found two 180 kpc HI tails as well as a faint optical tail displaced from the HI tails. They infer that the merger started 750 Myr ago and the two nuclei will completely merge in roughly 60 Myr. Evolutionary starburst models have shown that Arp 299 has been going through a recent episode of interaction-induced star formation over a span of about 15 Myr \citep{2000ApJ...532..845A}. Interferometric observations of $^{12}$CO show strong emission from the three major regions found in infrared images (Sargent $\&$ Scoville 1991; Aalto et al., 1997; Casoli et al., 1999)\nocite{1997ApJ...475L.107A,1999A&A...346..663C} indicating the presence of large amounts of molecular gas. 

Studying the physical conditions and distribution of the warm, dense gas associated with star formation will help in understanding the processes and timescales controlling the star formation in galaxy mergers. In this paper, we analyze Submillimeter Array (SMA) interferometric observations of Arp 299 in the $^{12}$CO J=2-1,  $^{13}$CO J=2-1 and $^{12}$CO J=3-2 lines first published in Wilson et al. (2008; hereafter Paper I)\nocite{2008ApJS..178..189W}.  In Section 2 we present the observations and the data reduction process including the short spacings correction. In $\S$ 3, we present and discuss our radiative transfer modelling results as well the CO-to-H$_{\rm{2}}$ conversion factor, $\alpha_{\rm{co}}$ for the extranuclear region. In $\S$ 4, we discuss the star formation rates and depletion times for each region of Arp 299.

\section{Observations and Data Reduction}

\subsection{Submillimeter Array} 

We have used the SMA (Ho et al. 2004)\nocite{2004ApJ...616L...1H} observations centered on $^{12}$CO J=2-1 and $^{12}$CO J=3-2 for Arp 299 first published in Paper I. The correlator was configured to have a spectral resolution of 0.8125 MHz corresponding to roughly 1.1 km s$^{-1}$ for $^{12}$CO J=2-1 and 0.7 km s$^{-1}$ for $^{12}$CO J=3-2. The $^{12}$CO J=2-1 transition was centered in the upper sideband and $^{13}$CO J=2-1 was observed simultaneously  in the lower sideband 10 GHz away. Since Arp 299 has sufficiently extended emission, a small mosaic of two pointings was used.  Further details of the observations are given in Paper I. 

\begin{figure}[h] 
\centering
$\begin{array}{@{\hspace{-0.2in}}c@{\hspace{0.1in}}c}
\includegraphics[scale=0.5]{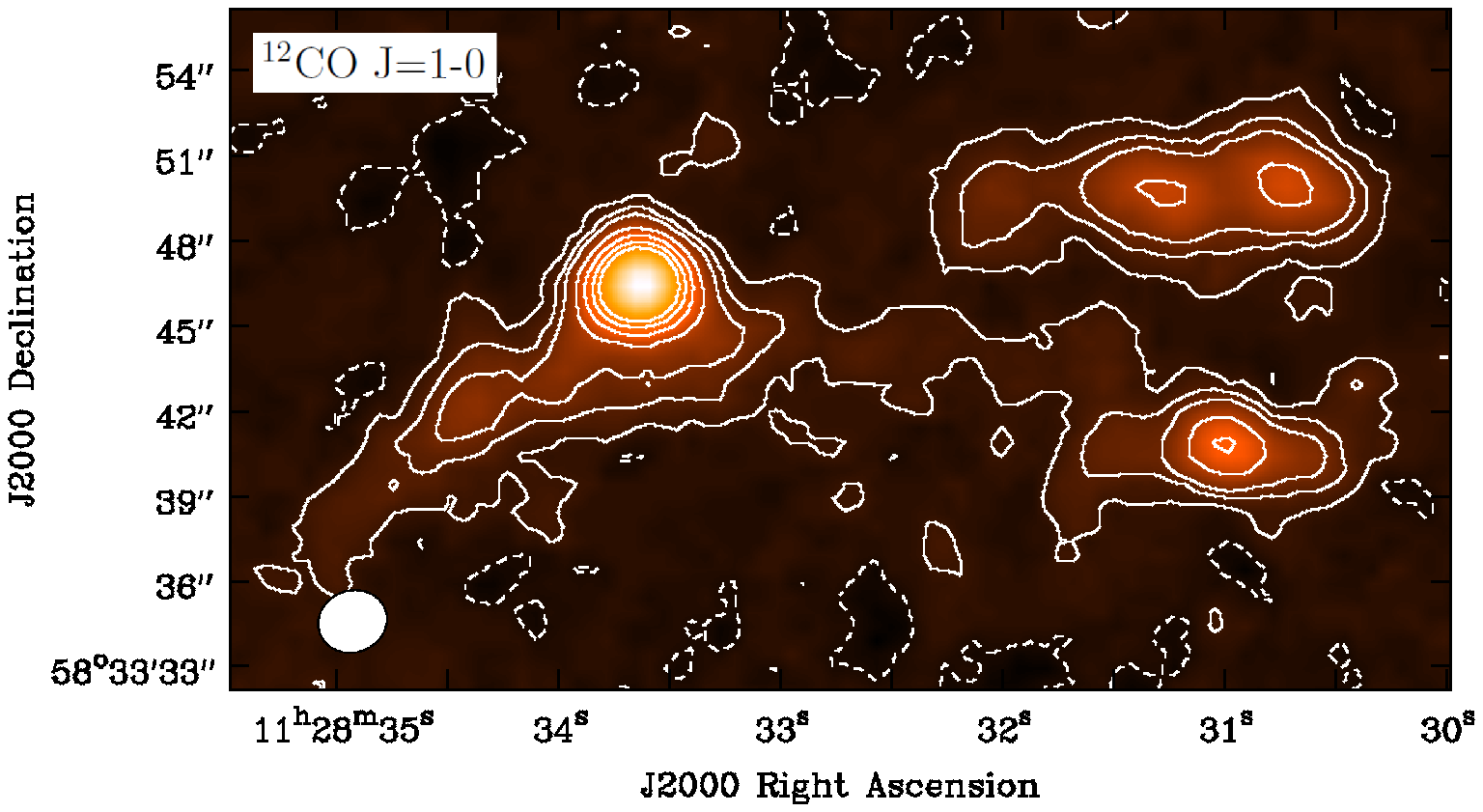} &\includegraphics[scale=0.5]{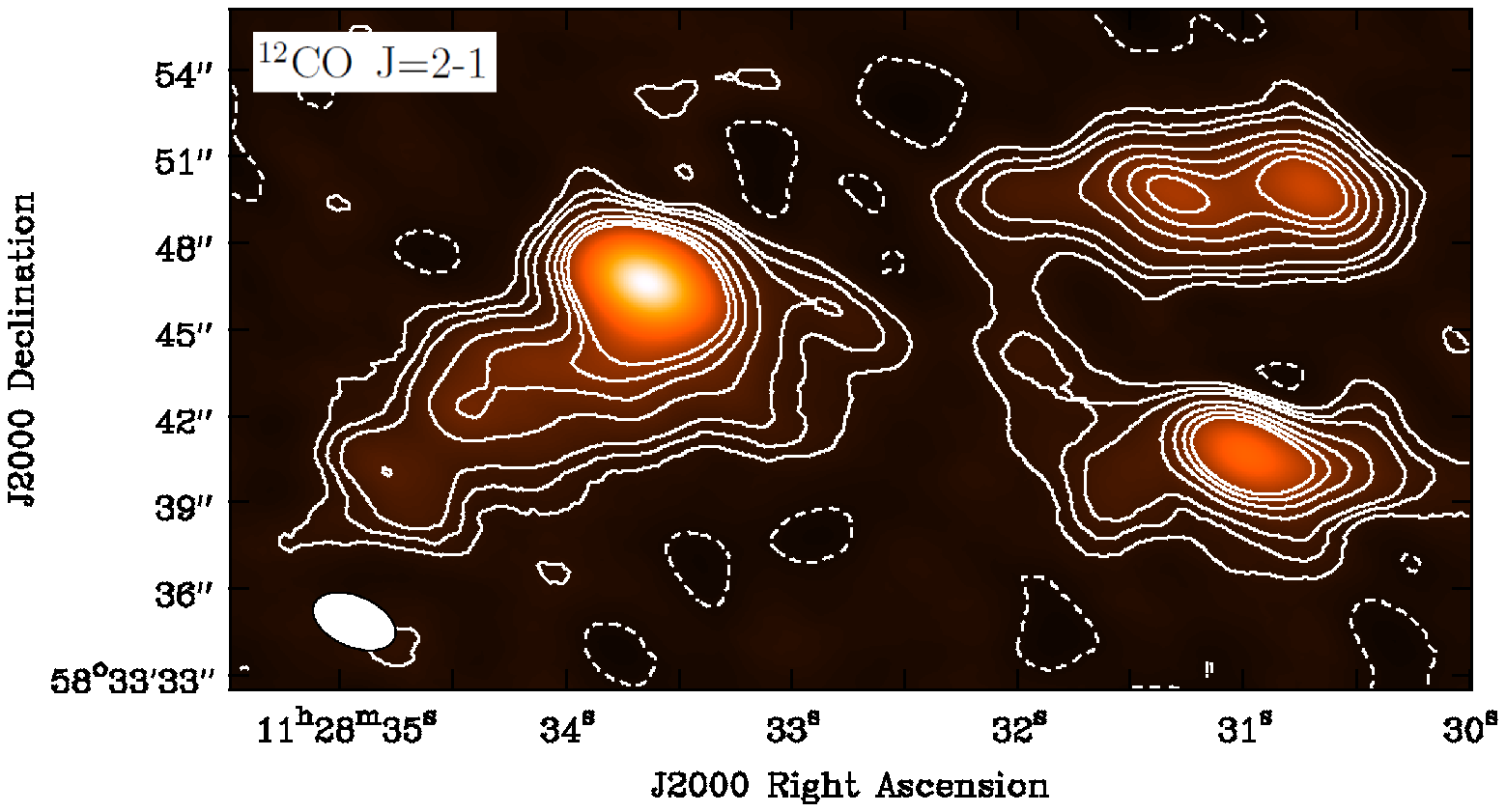} \\
\includegraphics[scale=0.5]{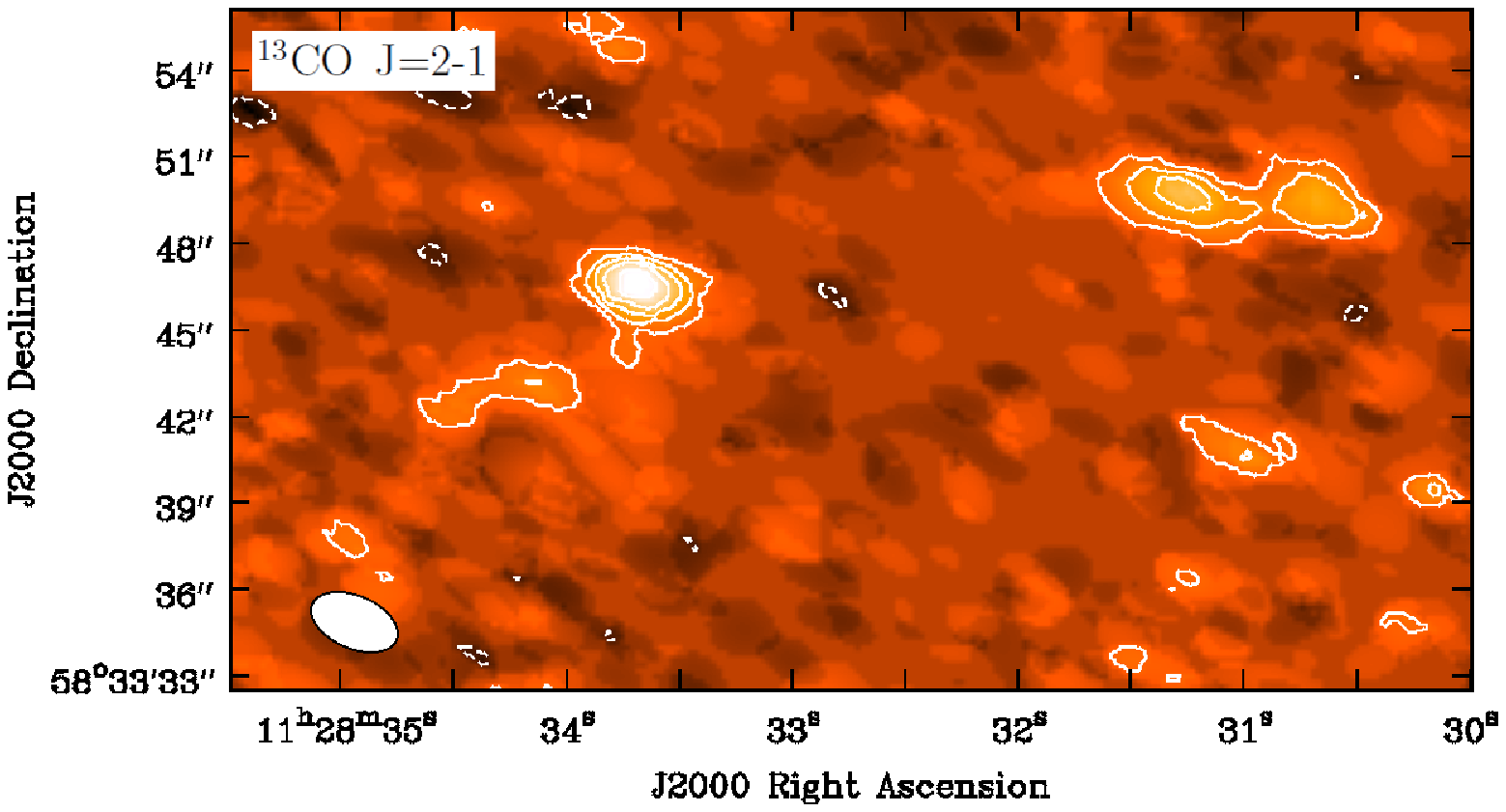} &\includegraphics[scale=0.5]{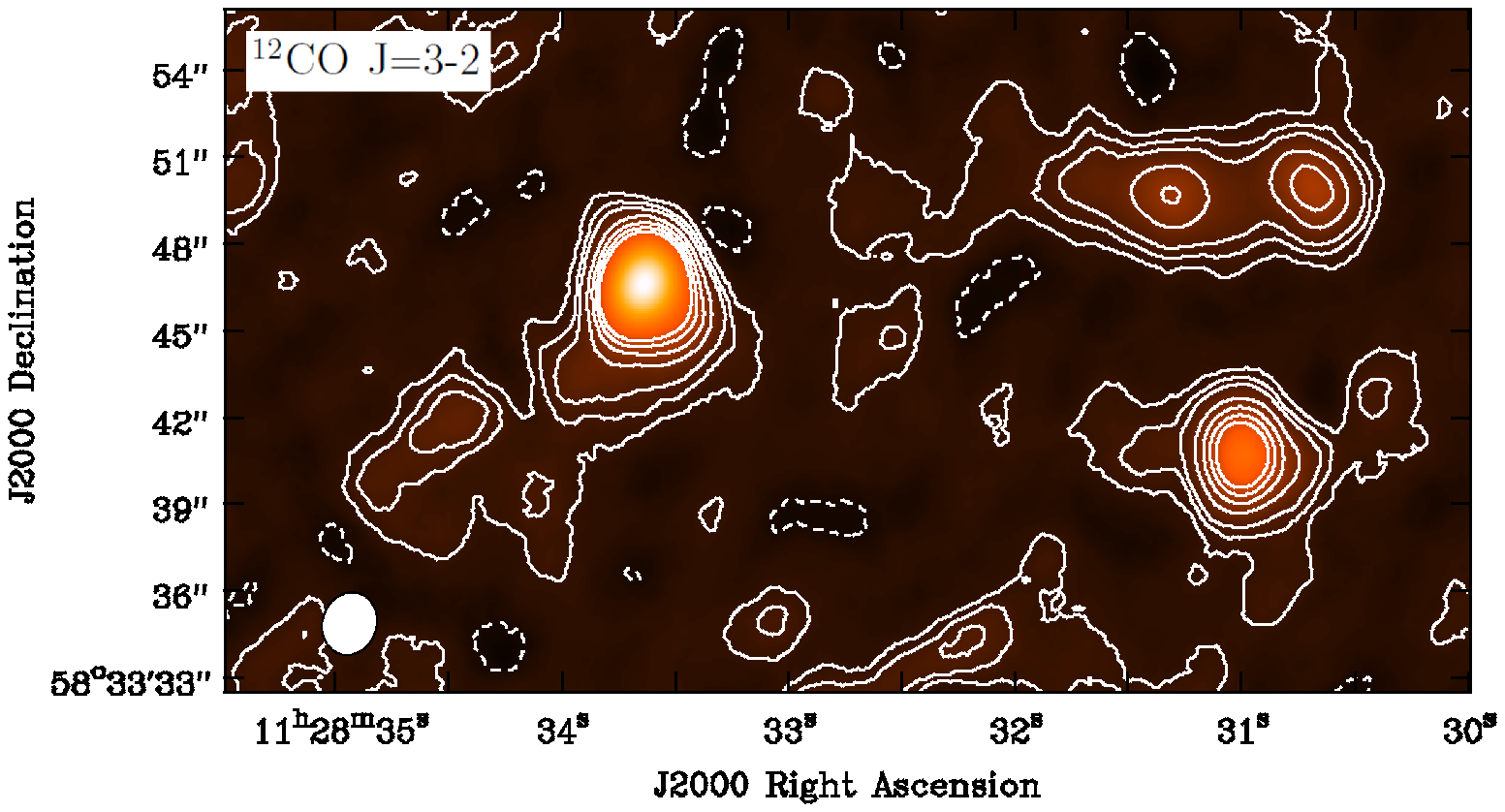}
\end{array}$
\caption[]{Integrated intensity maps of Arp 299. All SMA maps have been corrected for the primary beam. Dashed contours indicate negative flux: (\textit{top left}) $^{12}$CO J=1-0 OVRO map from data cube given by Aalto et al. (1997) with contours corresponding to -4, 2, 6, 10, 20, 30, 40, 50 $\times$ 0.9 Jy/beam km s$^{-1}$ (=1$\sigma$), (\textit{top right}) $^{12}$CO J=2-1 map with contours corresponding to -4, 2, 6, 10, 20, 30, 40, 50 $\times$ 1.7 Jy/beam km s$^{-1}$  (=1$\sigma$), (\textit{bottom left}) $^{13}$CO J=2-1 map with contours corresponding to -2, 2, 4, 6, 8, 10 $\times$ 1.4 Jy/beam km s$^{-1}$  (=1$\sigma$), (\textit{bottom right}) $^{12}$CO J=3-2 map with contours corresponding to  -4, 2, 6, 10, 20, 30, 40, 50 $\times$ 3.5 Jy/beam km s$^{-1}$  (=1$\sigma$). Note: The SMA maps from Paper I have been reprocessed using CASA.}
\label{SMAmaps}
\end{figure}

We start with the calibrated $uv$ data from Paper I. These data were converted to FITS files and then exported into CASA format for further processing and imaging.  The data were flagged to remove the first six and last six channels of each of the 24 chunks of the correlator. The data were also flagged to remove high amplitude value data points (removing less than 1$\%$ of the data). The $uv$ data sets were continuum subtracted using line-free channels. Clean data cubes with velocity resolutions of 20 km s$^{-1}$ and 50 km s$^{-1}$ were created using a robust  weighting scheme in order to achieve a good balance between resolution and sensitivity. Each data set was cleaned down to 2 times the rms noise in each velocity channel (Table 1). The beam was sufficiently clean and the emission is sufficiently compact that cleaning the entire inner quarter region produces good maps. However, for $^{12}$CO J=2-1 and J=3-2, we interactively placed clean boxes around emission regions in each channel. Integrated intensity maps were created with a 2$\sigma$ cutoff using velocity channels with emission from Arp 299 (Figure \ref{SMAmaps}). Finally, the integrated intensity maps were corrected for the primary beam. 

\subsection{Owens Valley Radio Observatory} 

We have obtained a clean data cube of $^{12}$CO J=1-0 observed using the Owens Valley Radio Observatory (OVRO) from S. Aalto first published in Aalto et al. (1997). Details of the observations and reduction are given in Aalto et al. (1997)\nocite{1997ApJ...475L.107A}. An integrated intensity map was created with a 2$\sigma$ cutoff using velocity channels with emission from Arp 299 (see Figure 1). The integrated intensity map was corrected for the primary beam. Adding this line to our analysis will help to better constrain the density and temperature of Arp 299. 

\subsection{James Clerk Maxwell Telescope} 

Arp 299 was observed with the James Clerk Maxwell Telescope (JCMT) on 2007 March 31 for $^{12}$CO J=3-2 (Program: M07AC11; PI: C.D. Wilson) and 2011 March 09 for $^{12}$CO J=2-1 (Program: M11AC06; PI: K. Sliwa). The JCMT was used to make a 2$\arcmin$ map with 6$\arcsec$ sampling using the array receiver HARP-B in the $^{12}$CO J=3-2 line with a 15$\arcsec$ beam and to make a 7$\arcsec$ sampled 1.5$\arcmin$ $\times$ 1$\arcmin$ map with the RxA receiver centered on the $^{12}$CO J=2-1 line with a 22$\arcsec$ beam. Further details of the $^{12}$CO J=3-2 data are given in Paper I. The raw data were made into a cube that spanned from -400 km s$^{-1}$ to 400 km s$^{-1}$ using the $Starlink$ software \citep{2008ASPC..394..650C} written for JCMT data. The cube was then baseline corrected using line free channels and a first order baseline. Several of the outer bad pixels were trimmed off and the cube was binned to a velocity resolution of 20 km s$^{-1}$.

We assume a main beam efficiency ($\eta_{\rm{mb}}$) of 0.6 for $^{12}$CO J=3-2 and 0.69 for $^{12}$CO J=2-1 in order to convert the intrinsic units of antenna temperature ($T_{\rm{A}}^{*}$) to main beam temperature ($T_{\rm{mb}}$ = $T_{\rm{A}}^{*}$/$\eta_{\rm{mb}}$) and then to Jy km s$^{-1}$ which gives us a scaling factor of 26.9 Jy K($T_{\rm{mb}}$)$^{-1}$ and 22.9 Jy K($T_{\rm{mb}}$)$^{-1}$ for $^{12}$CO J=3-2 and $^{12}$CO J=2-1, respectively. We thus assume that emission from Arp 299 fills the beam, unlike Paper I where Arp 299 was treated like a point source. 


\subsection{Short Spacing Correction} 

We use the JCMT maps to recover the short spacings of the SMA maps. We combine the two maps in the image plane using the $feather$ task in CASA. Both maps were binned to the same velocity resolution so that each velocity channel matches in each map. The feathering technique takes both maps, regrids the low resolution image to match the high resolution image, Fourier transforms them to the gridded $uv$-plane, sums up the gridded visibilities and then Fourier transforms the feathered map back into the image plane (Figure 2). Comparing the SMA only maps with the JCMT maps, the $^{12}$CO J=2-1 map was missing about 20$\%$ of the total flux and the $^{12}$CO J=3-2 map was missing about 70$\%$ of the total flux (Table 2). Note Paper I estimated only 47$\%$ missing flux in $^{12}$CO J=3-2 because they assumed Arp 299 is point-like in the K to Jy conversion of the single dish data. Casoli et al. (1999) have compared the OVRO $^{12}$CO J=1-0 map of Aalto et al. (1997) with an IRAM 30m single dish map and found that about 56$\%$ of the flux was missing. The missing flux in all the maps is an issue but we try to avoid the problem by using the peak integrated intensities for our analysis (see $\S$ 3.1).

\subsection{Line Ratio Maps} 

For our analysis, we degraded the resolution of all the interferometric maps to 3.6$\arcsec$ $\times$ 2.4$\arcsec$ in order to probe the molecular gas on similar physical scales. The elongated beam of the CO J=2-1 maps limited the best possible degraded resolution to 3.6$\arcsec$ $\times$ 2.4$\arcsec$. We applied a Gaussian taper weighting to the SMA $uv$ data to degrade the resolution. The resolution of the OVRO $^{12}$CO J=1-0 map was degraded by applying a Gaussian kernel to the data cube. New integrated intensity maps were made from these resolution matched cubes. We converted the intrinsic units of the CO integrated intensity maps from Jy/beam km s$^{-1}$ to K km s$^{-1}$. We then created $^{12}$CO $\frac{J=3-2}{J=2-1}$, $^{12}$CO $\frac{J=2-1}{J=1-0}$ and $\frac{\mathrm{^{12}CO}}{\mathrm{^{13}CO}}$ J=2-1 line ratio maps by dividing the appropriate maps with a cutoff at 2$\sigma$ for each map (Figure \ref{RatioMaps}). Note the line ratio values near the edges of each region are noiser because the uncertainty by definition approaches $\pm$50$\%$.

\section{Radiative Transfer Modelling}

\subsection{Physical Conditions in Arp 299}

We adopt the notation of \citet{1983ApJ...267..551G} to denote each region of Arp 299. The two nuclei are denoted as region A for IC 694 and region B for NGC 3690. We denote the ``disk" of region A as region A2. The strong emission to the north of region B, denoted as region C, can be broken up into two sub regions; the strongest peak to the west is denoted as C1 and the peak to the east is denoted as C2 (Figure 2). Note that our sub-region C2 is the region C2+C3 in Aalto et al. (1997).  

The line width and size of the CO emission of each region have been measured using the $^{12}$CO J=3-2 map which has the best angular resolution. The velocity line width FWHM of each region was fitted with a Gaussian and the results are presented in Table 3.  The diameter of each region was obtained by fitting to a two dimensional Gaussian. For region C, each sub-region, C1 and C2, was fitted to a two dimensional Gaussian because the entire region cannot be fitted well by a single Gaussian. The source sizes listed in Table 3 are deconvolved from the intrinsic $^{12}$CO J=3-2 beam size.

Using the line ratio maps, we obtain the line ratio values at the peak intensity pixel of each region found in the $^{12}$CO J=3-2 map. The line ratios and 1$\sigma$ values are presented in Table 4. The 1$\sigma$ values are calculated assuming a 20$\%$ calibration uncertainty for each map for the $^{12}$CO $\frac{J=3-2}{J=2-1}$ and $^{12}$CO $\frac{J=2-1}{J=1-0}$ line ratios and a 10$\%$ calibration uncertainty for each map for the $\frac{\mathrm{^{12}CO}}{\mathrm{^{13}CO}}$ J=2-1 line ratio since these maps were observed simultaneously. We see that including short spacings with the JCMT maps does not change the peak ratio values significantly except for region A2. We assume that the missing flux for $^{12}$CO J=1-0 and $^{13}$CO J=2-1 at the peak intensity value is insignificant for regions A, B, C1 and C2. The $^{12}$CO $\frac{J=2-1}{J=1-0}$ and $\frac{\mathrm{^{12}CO}}{\mathrm{^{13}CO}}$ J=2-1 line ratios for region A2 is more uncertain which may result in unreliable radiative transfer results.

To constrain the average temperature and density of each region, we use the radiative transfer code RADEX \citep{2007A&A...468..627V}. RADEX assumes a homogeneous medium making it a useful tool to constrain the physical conditions such as density and temperature using observational data. We created a 3D grid of models varying $N$($^{12}$CO) in steps of 0.1 $\times$ 10$^{18}$ cm$^{-2}$ from 10$^{17}$ to 10$^{20}$ cm$^{-2}$, $T_{\rm{kin}}$ in steps of 1 K from 10 - 1000 K and log($n$(H$_{\rm{2}}$)) in steps of 0.02 from 2 - 8. We adopt a $^{12}$CO to $^{13}$CO abundance ratio of 50. We used the line widths ($dV$) from Table 3 to convert N($^{12}$CO)/$dV$ to N($^{12}$CO). The RADEX solutions were plotted as $T_{\rm{kin}}$ vs. $n$(H$_{\rm{2}}$) with contours for each line ratio $\pm$1$\sigma$. The solutions are picked out where the three line ratios overlap within the $\pm$1$\sigma$ range by inspection by eye (Figure \ref{radexsolutions}). For regions A, A2 and B, a wide range of solutions exist. In particular, there are solutions with low ($<$ 30 K) temperature but density only constrained to be $>$ 10$^{4}$ cm$^{-3}$ and also solutions with lower densities but temperature only constrained to be $>$ 30 K. The solutions for regions A, A2, and B suggest that the pressure of the medium is P/k $\gtrsim$ 10$^{5}$ K cm$^{-3}$. Aalto et al. (1997) estimated a temperature of $>$ 50 K and a density of 10$^{4}$ -10$^{5}$ cm$^{-3}$ for region A using $^{12}$CO and $^{13}$CO J=1-0 and HCN J=1-0 line ratios. This is not consistent with our RADEX solutions but our analysis consists of more CO lines and we do not assume local thermal equilibrium (LTE). The HCN J=1-0 line traces molecular gas at a higher critical density than our $^{12}$CO J=3-2 line; however, the HCN lines may be affected by IR pumping through a 14 $\mu$m vibrational transition \citep{1995A&A...300..369A} and in extreme IR environments can be over-luminous with respect to the lines of other dense molecular gas tracers \citep{2006ApJ...640L.135G}. If the HCN J=1-0 line is over-luminous, radiative transfer solutions would show higher densities than the true value. For the sub-regions C1 and C2, we have better constrained solutions for both temperature and density (Table 5). For sub-region C2, the solutions show a slightly wider temperature, density and column density range than for region C1. The optical depth of the solutions for sub-regions C1 and C2 was found to be $\sim$0.5-1.5, $\sim$2.5-4 and $\sim$ 4 for $^{12}$CO J=1-0, J=2-1 and J=3-2, respectively. The low optical depth ($<$1) of $^{12}$CO J=1-0  line is likely responsible for the unusual ($>$ 1) 2-1/1-0 line ratios seen in Arp 299. The assumption of a homogeneous medium adds uncertainty to our solutions and the solution column densities may easily be off by a factor of 2 from the true values. We stress also that our solutions are averages over the beam. 

Our RADEX solutions for regions A, A2 and B show a wide range of solutions such as low ($<$ 30 K) temperature but density $>$ 10$^{4}$ cm$^{-3}$ and also solutions with lower densities but temperature only constrained to be $>$ 30 K.  One way this can be interpreted is a two gas component solution with a moderately dense warm ($T_{\rm{kin}}$ $>$ 30 K) gas component and a dense cold ($T_{\rm{kin}}$ $\sim$ 10-30 K) gas component. In Arp 220, \citet{2011ApJ...743...94R} have shown that the mid-J to high-J CO transition lines are tracing the warm/hot gas and the low-J CO transition lines are tracing the cold gas ($T \sim$ 50 K). The mass of the warm gas is 10$\%$ of the cold gas mass in Arp 220 (Rangwala et al. 2011). This result suggests that even our $^{12}$CO J=3-2 observations may be most sensitive to the dominant colder gas component that our RADEX solutions allow.  In order to trace the warm/hot gas component more definitively, we need to observe Arp 299 in higher $^{12}$CO transition lines. \citet{2010ApJS..188..447P} have used mid-infrared $Spitzer$ spectra of IC 694 and NGC 3690 to show that there is 5 $\times$ 10$^{8}$ $M_{\odot}$ of warm ($T_{\rm{kin}}$ $>$ 100K) H$_{2}$ gas. Since Arp 299 does have a warm/hot gas component up to $T_{\rm{kin}}$ $\sim$ 500 K, we would expect to see strong emission in higher J-level $^{12}$CO lines. The $Herschel$ FTS spectrum for Arp 299 indeed shows a prominent $^{12}$CO transition ladder with strong emission up to $^{12}$CO J=13-12 for regions A and B (M. R. P. Schirm; private communication)\nocite{schirm11}.

Recently, \cite{2012arXiv1202.1803P} have performed large velocity gradient (LVG) modelling of a single gas phase of Arp 299 using single dish CO transition lines. \cite{2012arXiv1202.1803P} used global $^{12}$CO $\frac{J=3-2}{J=1-0}$, $\frac{J=4-3}{J=1-0}$ and $\frac{\mathrm{^{12}CO}}{\mathrm{^{13}CO}}$ J=1-0 line ratios and found a best fit of $T_{\rm{kin}}$ = 30 K and $n$(H$_{\rm{2}}$) = 10$^{4}$ cm$^{-3}$, well within our RADEX solutions. Other solutions were shown to be possible with higher temperatures and low densities, similar to our own RADEX solutions for regions A, A2 and B.

\subsection{Arp 299 in Context: Comparison to M33}

LVG modelling has previously been performed on a sample of seven individual giant molecular clouds (GMCs) in the normal spiral galaxy M33 \citep{1997ApJ...483..210W}. The average of six clouds resulted in a solution of $T_{\rm{kin}}$ =10-20 K, $n$(H$_{\rm{2}}$) = 4-30 $\times$ 10$^{3}$ cm$^{-3}$ and $N$($^{12}$CO) = 3-6 $\times$ 10$^{17}$ cm$^{-2}$. This average solution resembles our RADEX solution for region C1 except for the beam-averaged column density which is significantly lower. Since the sizes of the clouds are known \citep{1990ApJ...363..435W} we can determine the average column density over the cloud itself using the filling factor of the cloud within the beam, $f$ = $D_{\rm{cloud}}^{2}/D_{\rm{beam}}^{2}$, where $D_{\rm{cloud}}$ is the diameter of the cloud and $D_{\rm{beam}}$ is the physical diameter of the beam (85 pc). Using the equation in \citet{1990ApJ...363..435W}
\begin{equation}
 V_{\mathrm{FWHM}} = 1.2 D_{\mathrm{pc}}^{0.5}
 \end{equation}
where $V_{\rm{FWHM}}$ is the line width FWHM of the CO line in km s$^{-1}$ and $D_{\rm{pc}}$ is the diameter of the cloud in pc and using the average $V_{\rm{FWHM}}$ for the six clouds given in Wilson et al. (1997)\nocite{1997ApJ...483..210W} (= 9.3 km s$^{-1}$), we get an average $D_{\rm{cloud}}$ of 60 pc. This gives us a filling factor of $f$ = 0.5 and a cloud-averaged column density of $N$($^{12}$CO) = 6-12 $\times$ 10$^{17}$ cm$^{-2}$ for the M33 clouds. Comparing the cloud-averaged column density value to the $^{12}$CO column density of region C, we see that the sub-region C1 and even C2 have at least 2-3 times the average column density of the clouds in M33. The striking thing is that this column density is achieved not just in an individual GMC but over a $\sim$1 kpc diameter region. This analysis implies that if region C (= C1 + C2) is made up of normal GMCs such as those in M33, the beam is filled with clouds spaced very close together such that the entire $\sim$1 kpc region has a structure similar to a single GMC and with at least 2-3 clouds along each line of sight. This unusual structure may help explain the short gas depletion time and high star formation rate observed in region C in particular and Arp 299 as a whole (see $\S$ 4). 

We can also calculate the average volume density of region C assuming a spheriodal geometry. The average radius in the plane of the sky of region C is $R$ $\sim$ 350 pc. However, a diameter of $\sim$ 700 pc  along the line of sight would be much greater than the typical disk scale height. From the column density comparison, we might expect to encounter a maximum of 3 GMCs of about 60 pc in diameter. If the space in between the clouds is negligible then the thickness of region C is about 200 pc. Using the molecular gas mass ($\S$ 4 and Table 6) and $<n$(H$_{\rm{2}}$)$>$ = $3M_{\rm{H_{2}}}/4\pi m_{\rm{H_{2}}}R^{3}$ we derive an average volume density of 130 cm$^{-3}$. We can compare this average volume density to the average value of the six clouds in M33. Using the equation given in Wilson $\&$ Scoville (1990)\nocite{1990ApJ...363..435W}
\begin{equation} \label{avgden}
<n(\mathrm{H_{2}})> = 200(D_{\mathrm{pc}} / 20 \mathrm{pc})^{-1}
\end{equation}
 and the average cloud diameter, we get an average volume density of about 70 cm$^{-3}$ for the clouds in M33.  This comparison implies that region C is much denser than a 60 pc cloud in M33, even though it covers a much larger spatial scale.
 
 It has been suggested previously that U/LIRGs have kiloparsec scale medium that is like a GMC as a whole (e.g. Downes $\&$ Solomon 1998). An alternative possibility that could reproduce the observed large column density is that U/LIRGs contain a medium where the mean density of individual clouds and clumps is much higher than typical GMCs (perhaps similar to the circumnuclear disk in the Galactic center i.e. Christopher et al. 2005)\nocite{2005ApJ...622..346C}. However this second situation would result in a radiative transfer solution volume density ($n_{\rm{RADEX}}$) much greater than that of the clouds in M33 ($n_{\rm{LVG}}$(M33)). For region C, we see that $n_{\rm{RADEX}}$ $\sim$ $n_{\rm{LVG}}$(M33) ruling out the alternative suggestion. \citet{2006PASJ...58..813I} have found that interferometric HCN/HCO+ J=1-0 line ratios of each region are less than 1 likely indicating that there is a lack of very dense gas, which also supports our results. \textit{This is the first direct evidence from a radiative transfer analysis using spatially resolved data of a kiloparsec scale structure having the properties of a GMC in a U/LIRG.}

\subsection{CO-to-H$_{2}$ Conversion Factor in Region C of Arp 299}

Since region C1 has a better constrained radiative transfer solution, we can examine both our assumed $^{12}$CO abundance ($x_{\rm{co}}$ = 3 $\times$ 10$^{-4}$) and also the value of the CO-to-H$_{2}$ conversion factor. We do this by comparing the RADEX solution's $^{12}$CO column density to the average H$_{2}$ column density within the beam derived from the $^{12}$CO intensity.  The CO luminosity of each region can be calculated using the equation
\begin{equation}\label{lumeqn}
{L^\prime \over {\rm K~km~s^{-1}~pc^2}} = 3.2546\times 10^7
({S_{\rm CO} \over {\rm Jy~km~s^{-1}}}) ({D_L \over {\rm Mpc}})^2
({\nu_0 \over {\rm GHz}})^{-2} (1+z)^{-1}
\end{equation}
where $S_{\rm{CO}}$ is the flux of the CO line in Jy km s$^{-1}$, $D_{\rm{L}}$ is the luminosity distance in Mpc, $\nu_{\rm{o}}$ is the rest frequency of the CO line in GHz and $z$ is the redshift (Paper I). Since the $^{12}$CO J=1-0 map does not include short spacings and we do not know how much flux is missing in each region, we use the short spacing corrected $^{12}$CO J=2-1 map to measure the gas masses of each region. We use an average ratio value of $^{12}$CO$\frac{J=2-1}{J=1-0}$ =1.4 (Table 4) to estimate the gas mass, which is then given by
\begin{equation}\label{massequation}
M(\mathrm{H_{2}}) = 0.57 L^{\prime}_{\mathrm{CO (2-1)}}
\end{equation}
where $L^{\prime}_{\rm{CO (2-1)}}$ is given by Equation \ref{lumeqn} and we assume $\alpha_{\rm{co}}$ = 0.8 $M_{\odot}$ (K km s$^{-1}$ pc$^{2}$)$^{-1}$ for the $^{12}$CO J=1-0 line \citep{1998ApJ...507..615D}. 

Using the short spacing corrected $^{12}$CO J=2-1 map at the degraded resolution, the peak intensity of region C1 is 170 Jy/beam km s$^{-1}$ or 2.2 $\times$ 10$^{8}$ K km s$^{-1}$ pc$^{2}$ which corresponds to a mass of $M_{\rm{H_{2}}}$ = 1.3 $\times$ 10$^{8}$ $M_{\odot}$ using Equation 4 . With a beam size of 3.6$\arcsec$ $\times$ 2.4$\arcsec$ this corresponds to an average H$_{\rm{2}}$ column density of 1.7 $\times$ 10$^{22}$ cm$^{-2}$. Using the $^{12}$CO column density obtained from the RADEX solution, we get a $^{12}$CO abundance of $x_{\rm{co}}$ = $N$($^{12}$CO)/$N$(H$_{\rm{2}}$) = 0.6 - 2.4 $\times$ 10$^{-4}$. This is 1.3 - 5 times lower than the abundance ratio of 3 $\times$ 10$^{-4}$ that is typically assumed \citep{2003ApJ...587..171W}. This difference may be interpreted in two ways. One possibility is that the abundance of $^{12}$CO to H$_{\rm{2}}$ is in fact lower in this region of Arp 299 than the average Galactic value. \citet{2007ApJ...667L.141R} have reported a metallicity that is near-solar for Arp 299 which would suggest that the abundance of $^{12}$CO to H$_{\rm{2}}$ should not differ greatly from the value that is typically used. If we use the updated work by \citet{1978ApJS...37..407D} for ``dark clouds" reported in \citet{1988ApJ...326..909M}, $N$(H$_{\rm{2}}$) = (4 $\pm$ 2)$\times$ 10$^{5}$$N$($^{13}$CO) and our assumed [$^{12}$CO]/[$^{13}$CO] of 50, we get an abundance of $x_{\rm{12CO}}$=(1.3 $\pm$ 0.7) $\times$ 10$^{-4}$. This value agrees well with our analysis. A second possibility is that the $^{12}$CO abundance in Arp 299 is the same as in the Galaxy but the value of the CO-to-H$_{\rm{2}}$ conversion factor, $\alpha_{\rm{co}}$, is smaller than 0.8 $M_{\odot}$(K km s$^{-1}$ pc$^{2}$)$^{-1}$. If we use the peak intensity value of the $^{12}$CO J=2-1 map and the mass within the beam derived from the column density solution of region C1 we get
\begin{equation}
\alpha_{\mathrm{CO (1-0)}}  = 0.4 \pm 0.3 (\frac{3 \times 10^{-4}}{x_{\mathrm{co}}}) \textrm{M$_{\odot}$(K km s$^{-1}$ pc$^{2}$)$^{-1}$}.
\end{equation}
To get $\alpha_{\rm{CO}}$ to agree with Downes $\&$ Solomon (1998), the $x_{\rm{co}}$ abundance factor must be $\sim$1.5 $\times$ 10$^{-4}$, lower than our assumed value by a factor of 2. There is no way of telling which scenario is right and it may well be a combination of the two. \cite{2012arXiv1202.1803P} have also found a CO-to-H$_{\rm{2}}$ factor over their global LVG solutions to be 0.35 - 0.42 $M$$_{\odot}$ (K km s$^{-1}$ pc$^{2}$)$^{-1}$. This range of values agrees well with the value we estimated for region C1 using the RADEX solutions. Note that our result is independent of the method used by Downes $\&$ Solomon (1998).

We can attempt to place a second limit on the CO-to-H$_{\rm{2}}$ conversion factor using the dynamical mass of region C (=C1+C2). The dynamical mass (Table 3) can be estimated using the equation from \citet{1990ApJ...363..435W} 
\begin{equation}\label{dynmass}
M_{\mathrm{dyn}} = 99\Delta V_{\mathrm{FWHM}}^{2} D\mbox{(pc)} M_{\odot}
\end{equation}
where $\Delta V_{\rm{FWHM}}$ is the CO line FWHM in km s$^{-1}$ and $D$(pc) is the diameter of the region in pc. Note equation \ref{dynmass} is valid only for regions that are gravitationally bound. If we assume that the dynamical mass is entirely due to molecular gas, then $\alpha_{\rm{CO}}$ = $M$$_{\rm{dyn}}$/$L$$_{\rm{CO}}^{\prime}$. Using the total flux for region C from the short spacing corrected $^{12}$CO J=2-1 map and a $^{12}$CO 2-1/1-0 average ratio of 1.4 (Table 4), we find that $\alpha_{\rm{CO}}$ = 1.4 $M$$_{\odot}$ (K km s$^{-1}$ pc$^{2}$)$^{-1}$. However, the dynamical mass is likely greater than the true molecular gas mass as stars may contribute some mass and the region as a whole may not be gravitationally bound. Thus, this value of $\alpha_{\rm{CO}}$ is strictly speaking an upper limit.


 \section{Star Formation Rates and Depletion Times}
 
We estimate the far-infrared luminosity of each region using $L_{\rm{FIR}}$ = 5.5 $\times$ 10$^{11}$ $L_{\odot}$ \citep{2003AJ....126.1607S} for Arp 299 and the fractions of the total luminosity for each region from Alonso-Herrero et al. (2000)\nocite{2000ApJ...532..845A} based on ground-based mid-infrared observations. From the relative fluxes at 10, 12, 20 and 30 $\mu$m through a 6$\arcsec$ diameter, Alonso-Herrero et al. (2000) estimated that region A has about 50$\%$, region B has about 27$\%$ and region C has about 13$\%$ of the total infrared luminosity. The final 10$\%$ is believed to be associated with H II regions. The star formation rate can be estimated from the infrared luminosity (8-1000 $\mu$m) using the equation from Kennicutt (1998)\nocite{1998ARA&A..36..189K}, $\dot{M}_{\rm{SFR}}\mbox{($M_{\odot}$ yr$^{-1}$)}= 4.5 \times 10^{-44} L_{\rm{IR}}\mbox{(erg s$^{-1}$)}$. Kennicutt (1998) used a Salpeter \citep{1955ApJ...121..161S} initial mass function (IMF) to estimate the star formation rate; a double power law IMF with a slope of -1.3 in the range of 0.1-0.5 $M$$_{\odot}$ and a slope of -2.3 in the range of 0.5-120 $M$$_{\odot}$ is more commonly used (i.e. Calzetti et al. 2007)\nocite{2007ApJ...666..870C}. The star formation rate is 1.59 times lower with the double power law IMF than with the Salpeter IMF. To minimize any contribution to the infrared luminosity from any AGNs present in Arp 299, we adopt the method of \citet{2010MNRAS.407.2091G}, who scale the far-infrared luminosity by a factor of 1.3 to estimate the total infrared (8-1000 $\mu$m) luminosity associated with star formation. This method avoids using the directly measured mid-infrared luminosity that is more likely to be contaminated by an AGN. The correction factor of 1.3 is an average $L_{\rm{IR}}$/$L_{\rm{FIR}}$ in nearby star-forming galaxies \citep{2008A&A...479..703G}. With these changes, the star formation rate becomes
\begin{equation}
\dot{M}_{\mathrm{SFR}}\mbox{($M_{\odot}$ yr$^{-1}$)}= 1.3 \times 10^{-10} L_{\mathrm{FIR}}\mbox{($L_{\odot}$)}.
\end{equation}
Table 6 summarizes the derived properties of Arp 299. 

Using the total molecular gas mass and star formation rates of each region, we can estimate the amount of time it will take for all the molecular gas to be depleted by star formation using
\begin{equation}
t_{\mathrm{depl}} = \frac{1.36M(\mathrm{H_{2}})}{\dot{M}_{\mathrm{SFR}}}
\end{equation}
where the factor 1.36 is to account for the helium gas. The depletion times of each region (Table 6) are very short compared to normal spiral galaxies \citep{2008AJ....136.2782L}. The timescales are more similar to the lifetime of GMCs in the Large Magellenic Cloud (20-30 Myr; Kawamura et al., 2009)\nocite{2009ApJS..184....1K}. \citet{2010ApJ...714L.118D} and \citet{2010MNRAS.407.2091G} studied the star formation laws in disk galaxies and starbursts at low and high redshifts. They provide evidence that disks and starbursts occupy distinct regions in the molecular gas mass ($M_{\rm{H_{2}}}$) versus star formation rate (or $L_{\rm{IR}}$) plane. Starbursts are seen to have 10 times higher $L_{\rm{IR}}$ at fixed $M_{\rm{H_{2}}}$. U/LIRGs have depletion times on the order of 10$^{7}$ years while spiral galaxies have depletion times in the range of 0.6 - 2.4 Gyr \citep{2010ApJ...714L.118D}. 

To increase the gas depletion times in Arp 299 to values similar to normal spiral galaxies, either the molecular gas mass is underestimated or the star formation rate is overestimated. The molecular gas mass must be increased by 2 orders of magnitude to achieve similar depletion times to those seen in normal spiral galaxies. However, even if we take the dynamical masses to be entirely molecular gas, we do not get close to increasing the gas mass by this amount. If the far-infrared luminosity is heavily contaminated by an AGN then the star formation rates could be overestimated. If we take the extreme case that an AGN contributes 80$\%$ to the total infrared luminosity, the largest contribution found for any U/LIRG by \citet{1998ApJ...498..579G}, the depletion timescales in the two nuclei would only be increased by a factor of 5. In either case, taking the molecular gas mass or the star formation rates to the extreme limits does not give us depletion timescales similar to those of normal spiral galaxies.

The global gas depletion time of Arp 299 suggests that all of the molecular gas will be used up in star formation in about half of the merger time ($\sim$ 60 Myr; Hibbard $\&$ Yun 1999). If the molecular gas is not replenished over time, Arp 299 may not reach the ULIRG phase. However, Hibbard $\&$ Yun (1999) have detected a tidal tail containing 3.3 $\times$ 10$^{9}$ M$_{\odot}$ of HI. If the gas inflow rate is sufficiently fast then there could be replenishment of molecular gas for further star formation. 
 
\section{Conclusions}

We have analyzed the interferometric CO observations of the local LIRG Arp 299 first published in Wilson et al. (2008). CO line ratios have been used to constrain the physical conditions of each region of Arp 299 using the radiative transfer code RADEX. Our RADEX solutions for regions A, A2 and B show two situations, solutions with low ($<$ 30 K) temperature but density only constrained to be $>$ 10$^{4}$ cm$^{-3}$ and also solutions with lower densities but temperature only constrained to be $>$ 30 K.  This analysis allows for the possibility of a two component gas, but higher J-level $^{12}$CO transition lines are required to trace the warm gas conclusively. 

The sub-region C1 corresponding to the peak intensity overlap region of Arp 299 shows a better-constrained RADEX solution with $T_{\rm{kin}}$ $\sim$ 10-50 K, $n$(H$_{\rm{2}}$) $\sim$ 1 - 30 $\times$ 10$^{3}$ cm$^{-3}$ and $N$($^{12}$CO) $\sim$ 1-4 $\times$ 10$^{18}$ cm$^{-2}$. This solution is comparable to that of GMCs in the normal spiral galaxy M33. The column densities of the clouds of M33 are at least 2-3 times lower than the column density for region C (= C1 + C2) which may indicate that clouds are spaced close together in region C. We estimate the CO-to-H$_{\rm{2}}$ factor, $\alpha_{\rm{co}}$ to be 0.4 $\pm$ 0.3 (3 $\times$ 10$^{-4}$/ x$_{\rm{co}}$) $M_{\odot}$ (K km s$^{-1}$ pc$^{2}$)$^{-1}$ for region C1, which agrees well with Downes $\&$ Solomon (1998) for more compact and luminous ULIRGs. Using the dynamical mass of region C (= C1 + C2), an upper limit to $\alpha_{\rm{CO}}$ was determined to be 1.4 $M$$_{\odot}$ (K km s$^{-1}$ pc$^{2}$)$^{-1}$

We have used the CO observations to calculate the gas and dynamical masses of each region and infrared data to derive the star formation rate of each region. We find that the gas depletion times of each region (20-60 Myr) are about 2 orders of magnitude lower than those for normal spiral galaxies and resemble the 30 Myr depletion timescales of GMCs. These short depletion timescales may be explained by a higher concentration of dense gas than in normal spiral galaxies. If the molecular gas is not replenished over time, Arp 299 will run out of fuel for star formation in about half the time it will take for the nuclei to finally merge.

\acknowledgments

The Submillimeter Array is a joint project between the Smithsonian Astrophysical Observatory and the Academia Sinica Institute of Astronomy and Astrophysics and is funded by the Smithsonian Institution and the Academia Sinica.
The James Clerk Maxwell Telescope is operated by The Joint Astronomy Centre on behalf of the Particle Physics and Astronomy Research Council 
of the United Kingdom, the Netherlands Organisation for Scientific Research, and the National Research Council of Canada.
We thank the anonymous referee for a very useful referee report and S. Aalto for giving us the OVRO CO J=1-0 map. C.D.W. acknowledges support by  the Natural Science and Engineering Research Council of Canada (NSERC).


\begin{figure}[h] 
\centering
\includegraphics[scale=0.75]{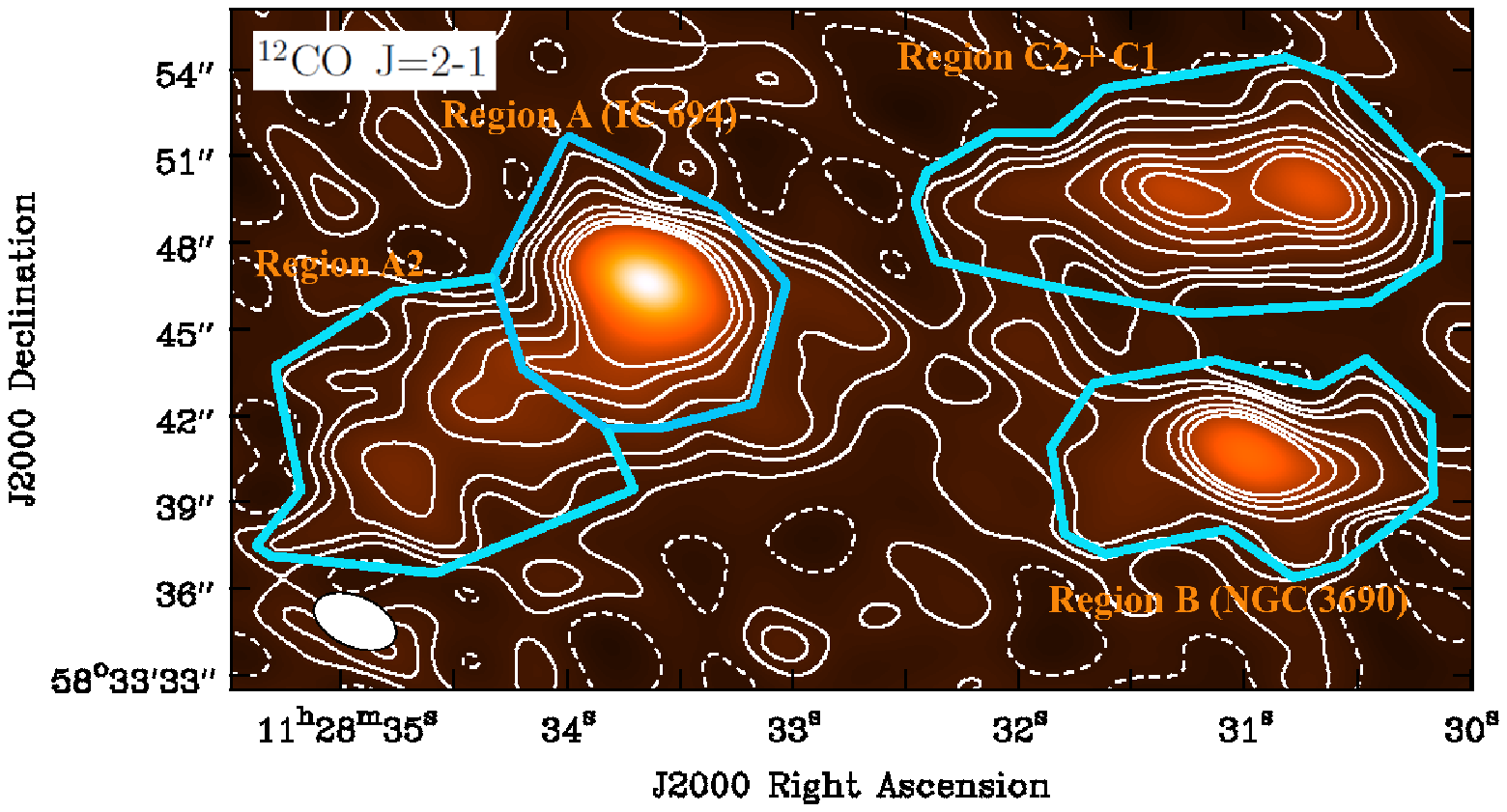}
\includegraphics[scale=0.75]{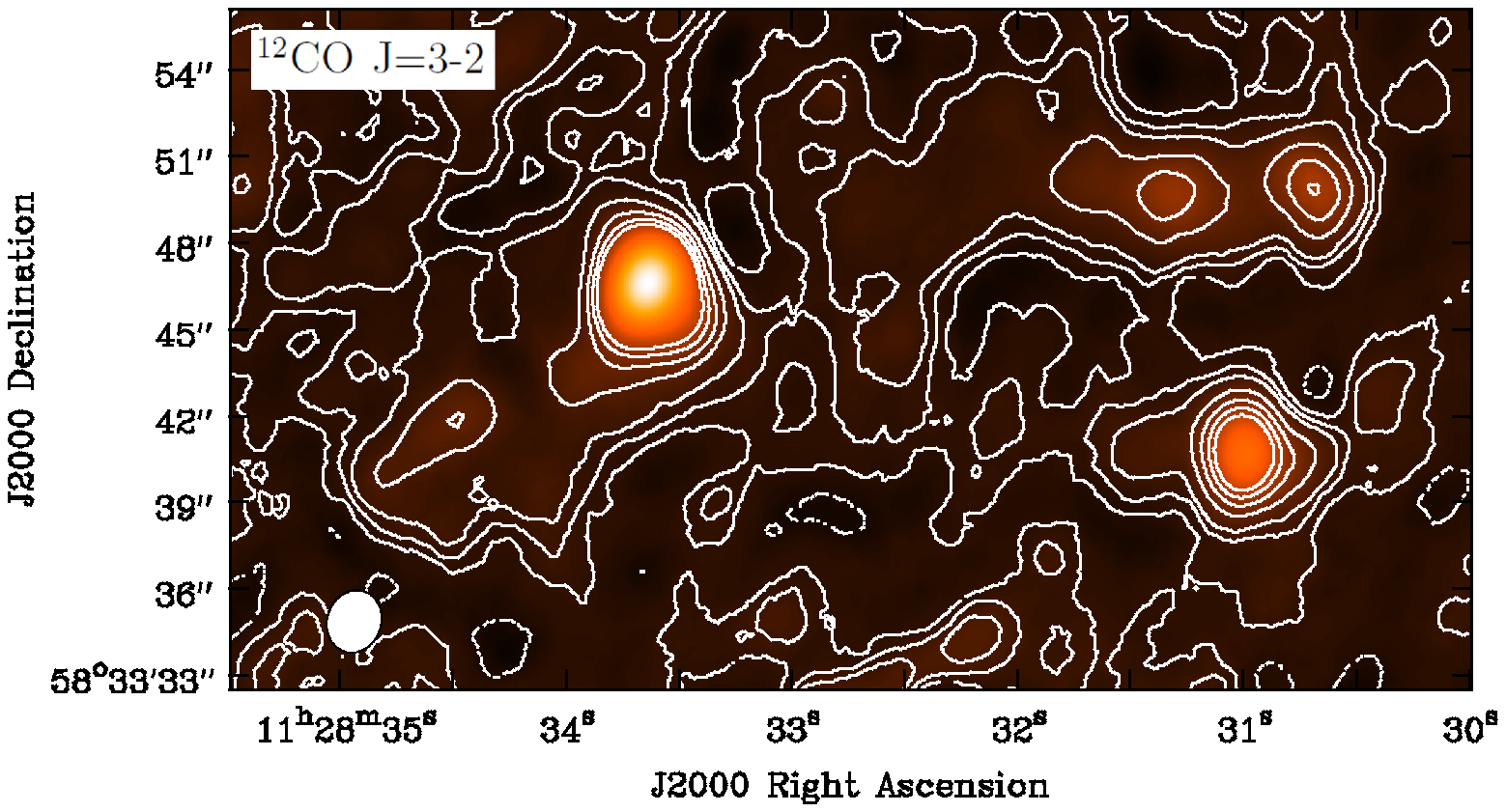}
\caption[Feathered SMA maps]{Feathered (short spacing corrected) SMA maps: ($top$) $^{12}$CO J=2-1 map with contour levels as in Figure \ref{SMAmaps}. The light blue lines indicate the area where the flux was integrated for each region (see Table 6); ($bottom$) $^{12}$CO J=3-2 map with contour levels as in Figure \ref{SMAmaps}. }
\label{feathered}
\end{figure}

\begin{figure}[h] 
\centering
$\begin{array}{c@{\hspace{0.5in}}c}
\includegraphics[scale=0.35]{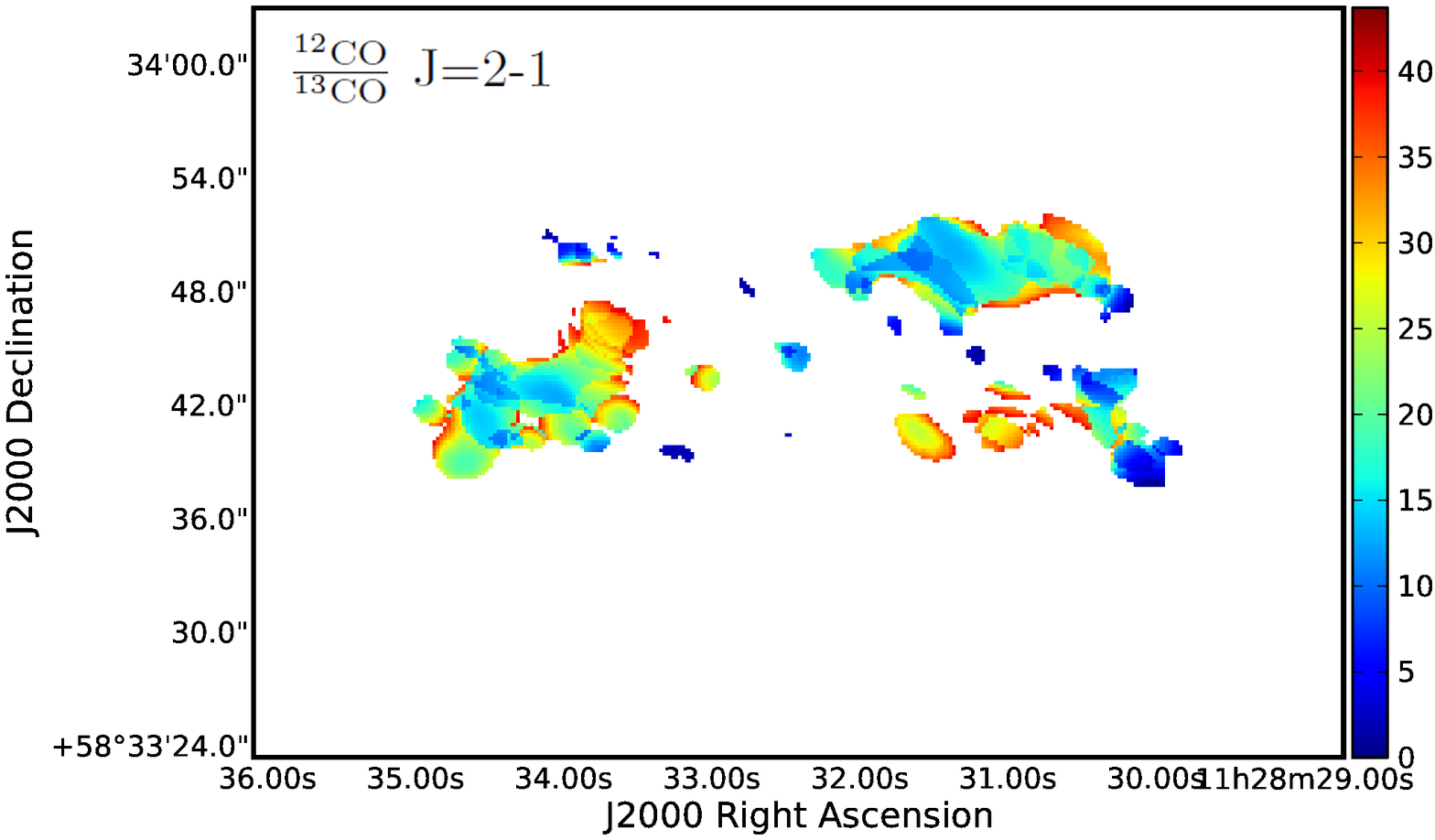} & \includegraphics[scale=0.35]{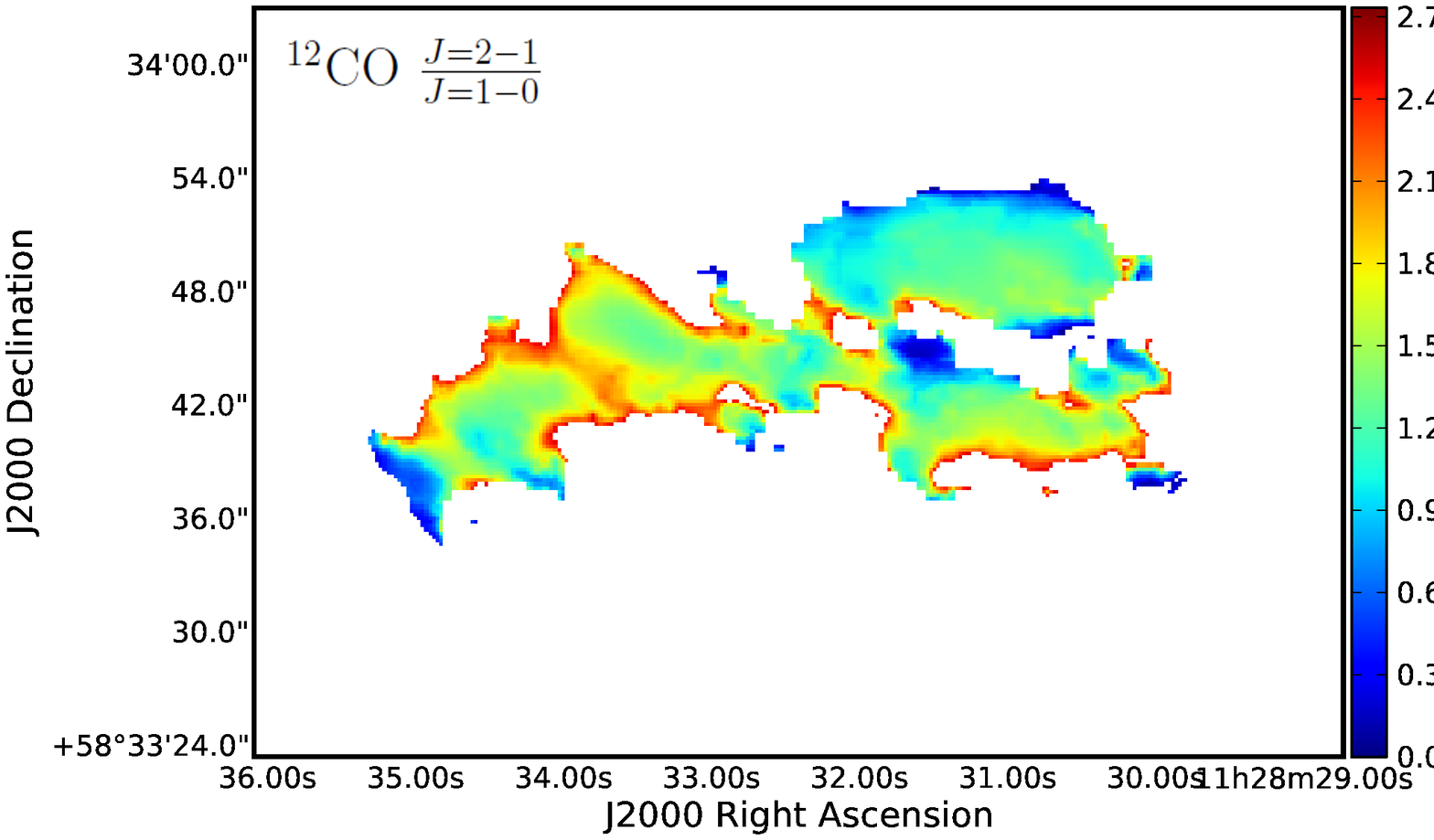} \\
\includegraphics[scale=0.35]{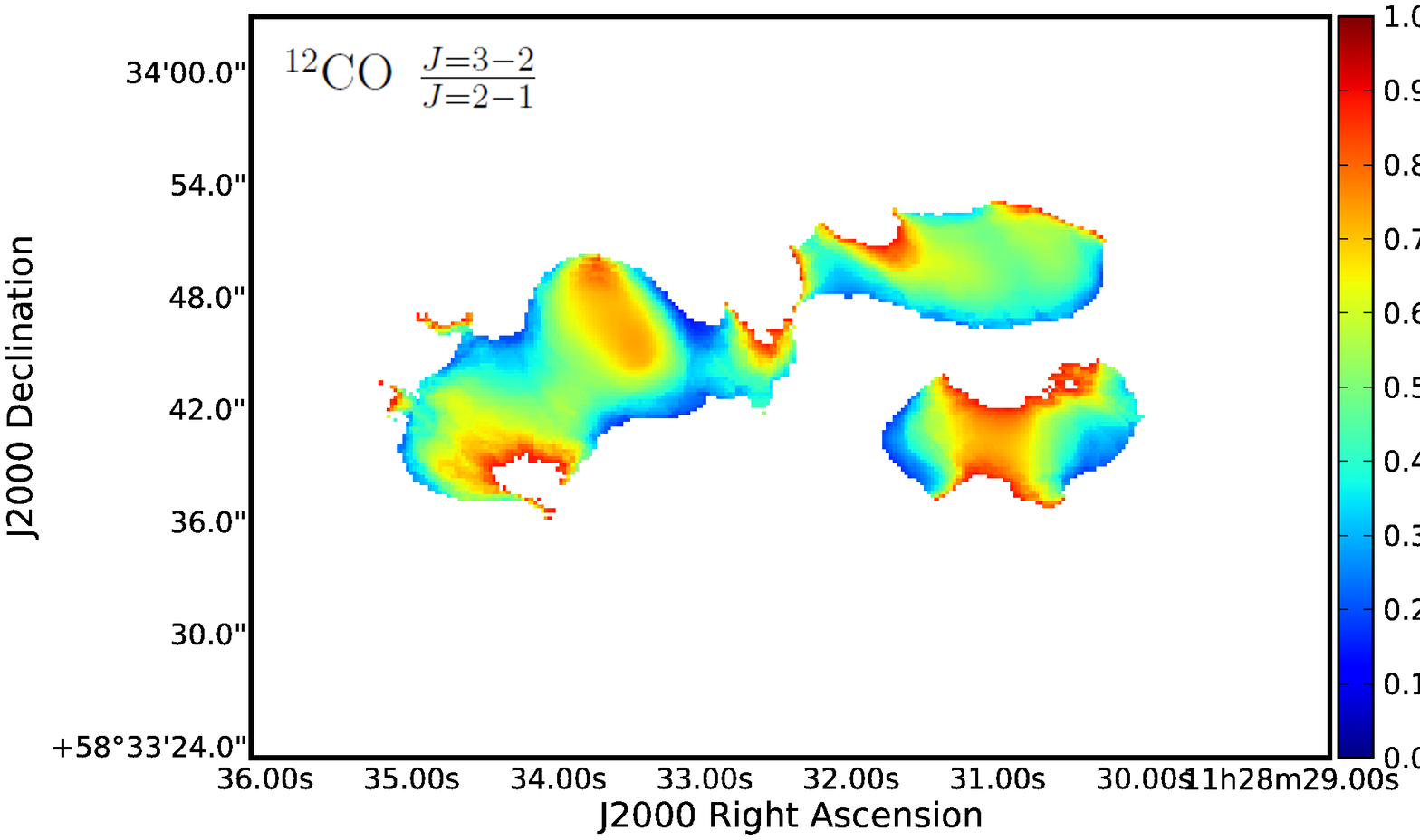} & \includegraphics[scale=0.35]{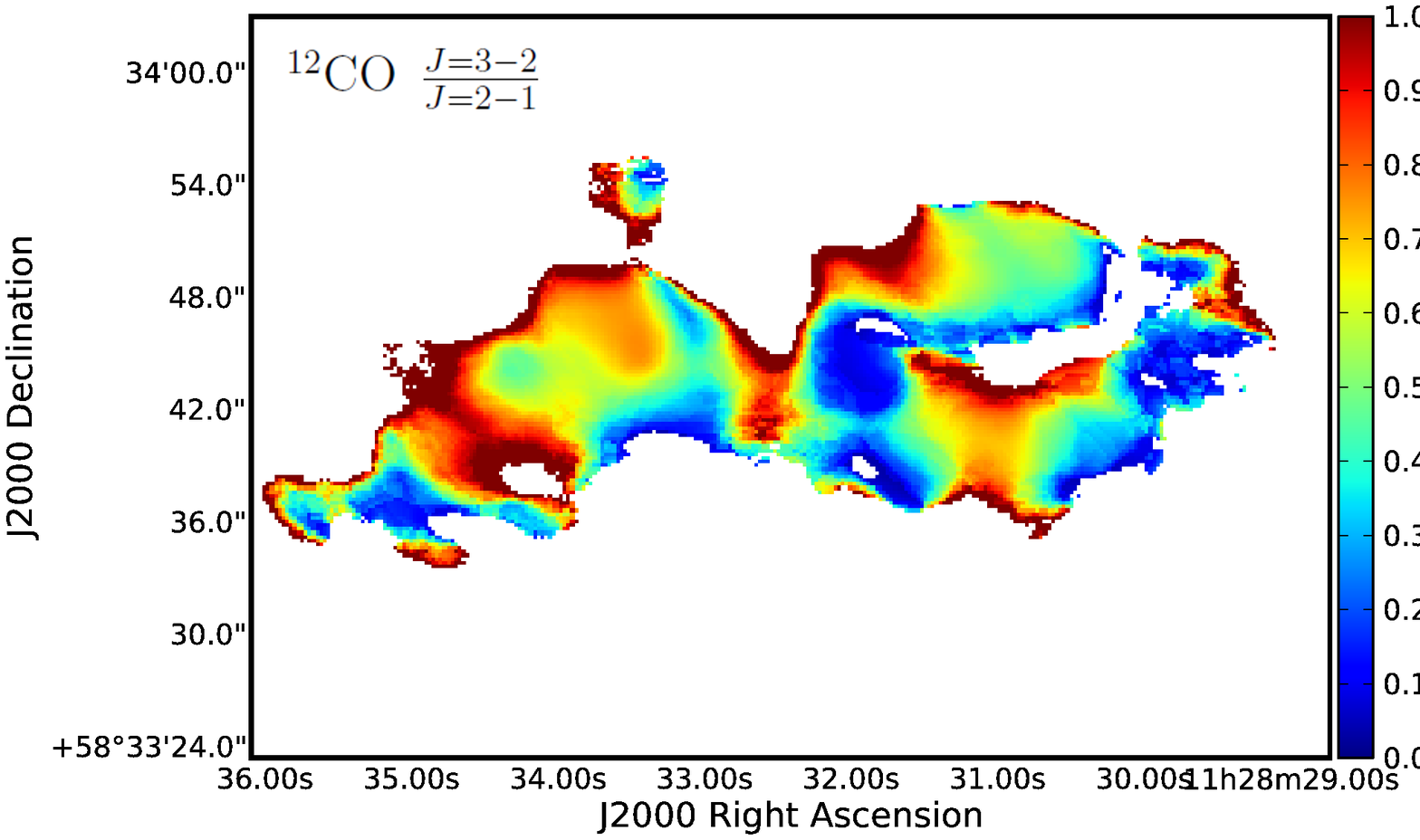} \\
\end{array}$
\caption[Line ratio maps]{Line ratio maps: (\textit{Top Left}) $\frac{\mathrm{^{12}CO}}{\mathrm{^{13}CO}}$ J=2-1; (\textit{Top Right}): $^{12}$CO $\frac{J=2-1}{J=1-0}$ ; (\textit{Bottom Left}): SMA only $^{12}$CO $\frac{J=3-2}{J=2-1}$; (\textit{Bottom right}): Feathered $^{12}$CO $\frac{J=3-2}{J=2-1}$ ratio map. Only emission that is $>$ 2$\sigma$ in both maps is included. }
\label{RatioMaps}
\end{figure}

\begin{figure}[h] 
\centering
$\begin{array}{c@{\hspace{0.05in}}c@{\hspace{0.05in}}c}
\includegraphics[scale=0.7]{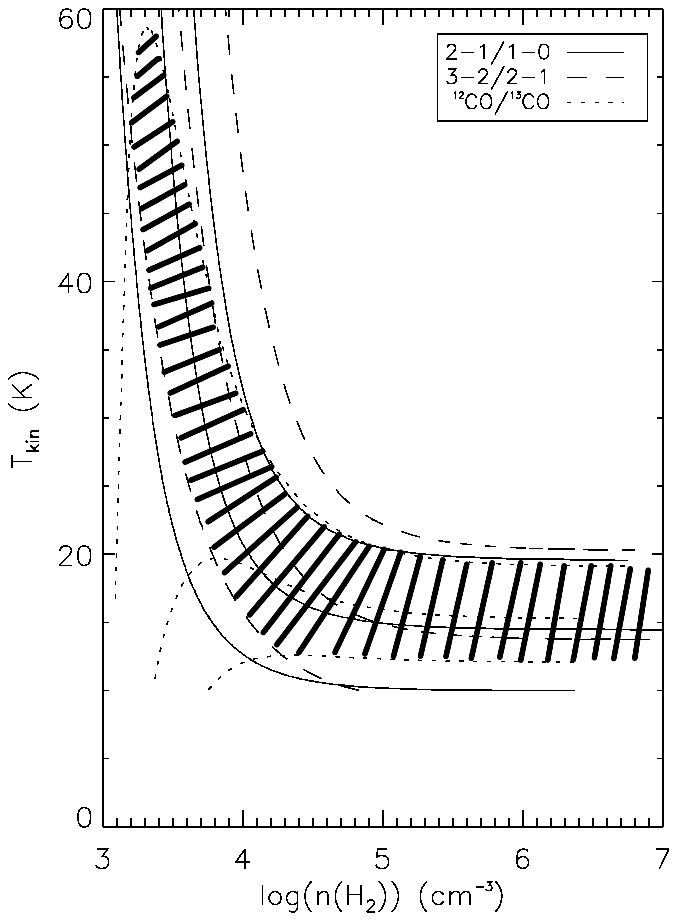} & \includegraphics[scale=0.7]{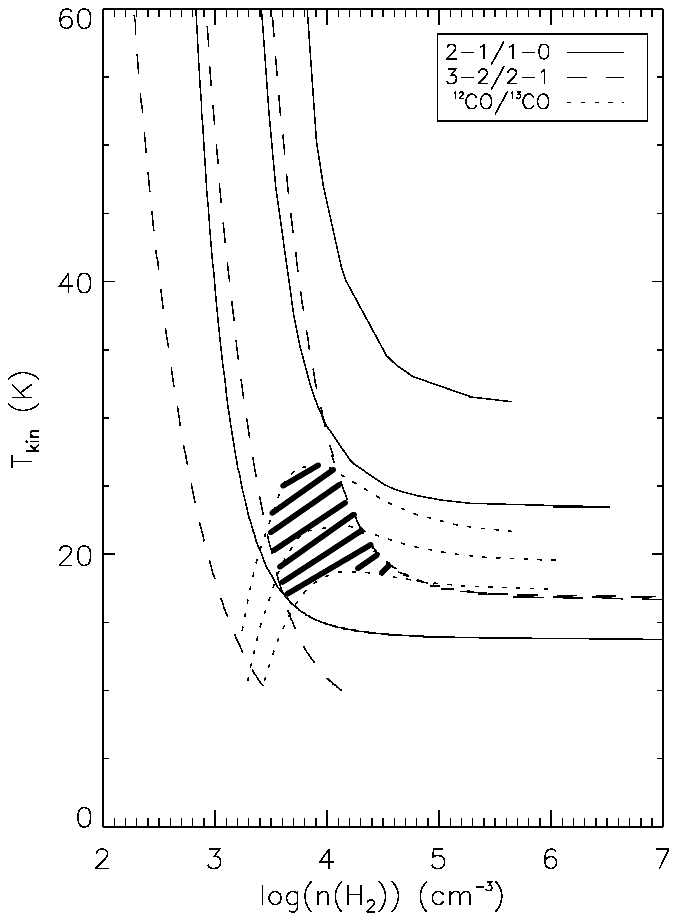} & \includegraphics[scale=0.7]{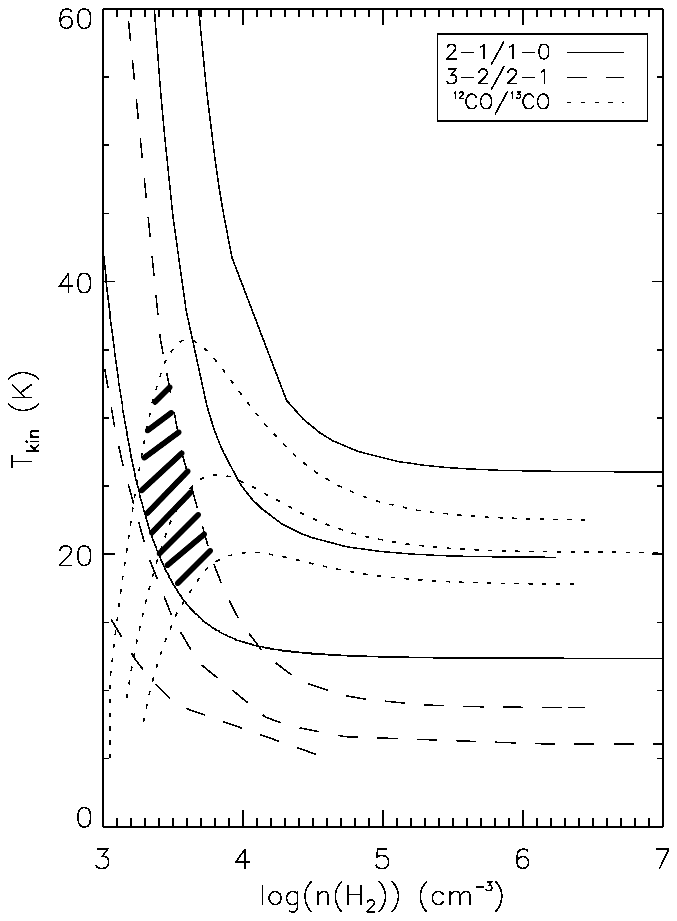} \\
\end{array}$
$\begin{array}{c@{\hspace{0.05in}}c}
\includegraphics[scale=0.7]{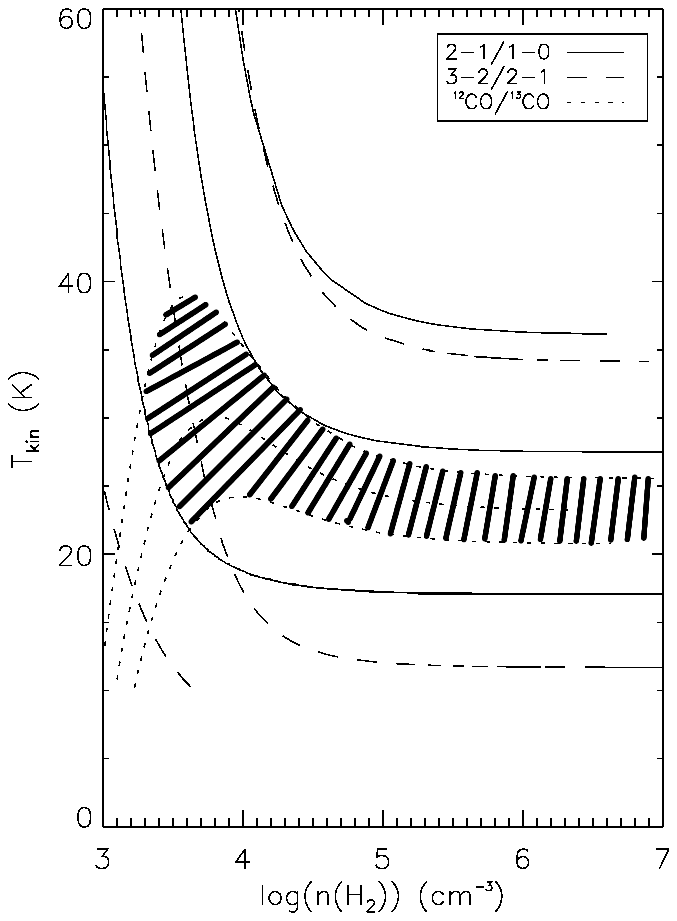} & \includegraphics[scale=0.7]{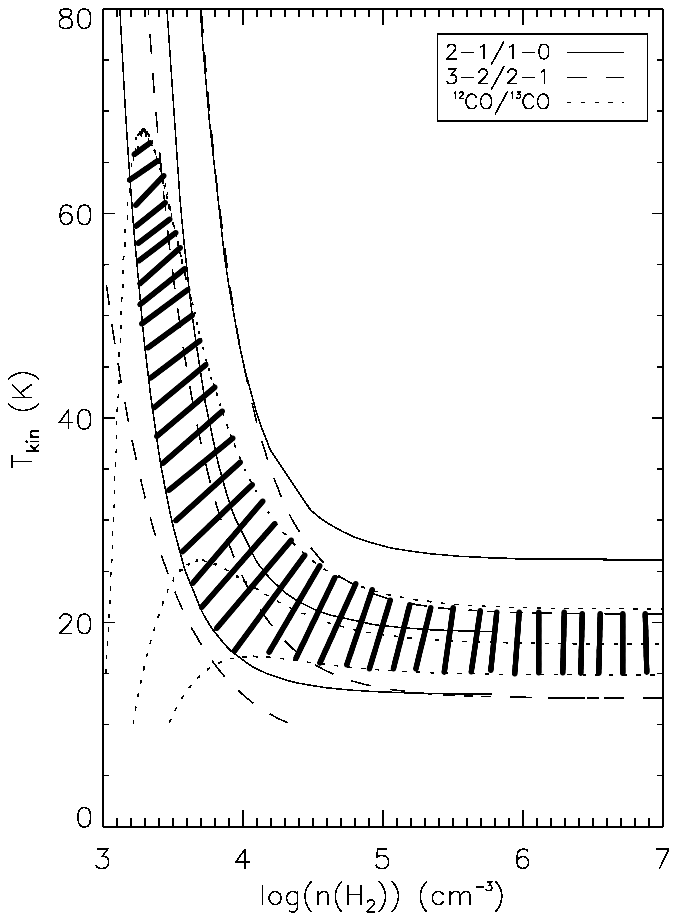} 
\end{array}$
\caption[Sample RADEX plots of each region of Arp 299]{Sample RADEX plots of each region of Arp 299: (\textit{top left}) Region A for $N$($^{12}$CO) = 2 $\times$ 10$^{18}$ cm $^{-2}$, (\textit{top middle}) Sub-region C2 for $N$($^{12}$CO) = 3 $\times$ 10$^{18}$ cm $^{-2}$, (\textit{top left}) Sub-region C1 for $N$($^{12}$CO) = 2 $\times$ 10$^{18}$ cm $^{-2}$, (\textit{bottom left}) Region A2 for $N$($^{12}$CO) = 4 $\times$ 10$^{18}$ cm $^{-2}$ and (\textit{bottom right}) Region B for $N$($^{12}$CO) = 2 $\times$ 10$^{18}$ cm $^{-2}$. The thick solid lines denote solutions. The entire range of solutions is not shown.}
\label{radexsolutions}
\end{figure}


\begin{deluxetable}{cccc} 
\tablecolumns{4}
\tablewidth{0pt}
\tablecaption{Interferometric Data For Arp 299}
\label{smasum}
\tablehead{\colhead{CO Transition Line} & \colhead{Total Flux\tablenotemark{a}} & \colhead{Beam} & \colhead{rms}\\ \colhead{} & \colhead{(Jy km s$^{-1}$)} & \colhead{($\arcsec$)} & \colhead{(mJy/beam)\tablenotemark{b}} }
\startdata
$^{12}$CO J=1-0\tablenotemark{c}	&377$\pm$ 9		&2.5 $\times$ 2.2 &9\\
$^{12}$CO J=2-1				&1870 $\pm$ 30	&3.0 $\times$ 1.8 &17\\
$^{13}$CO J=2-1				& 40 $\pm$ 2		&3.2 $\times$ 1.8 &10\\
$^{12}$CO J=3-2				&2820 $\pm$ 40	&2.2 $\times$ 1.9 &33\\
\enddata
\tablenotetext{a}{measurement uncertainty only; calibration uncertainty is 20$\%$ (Paper I).  }
\tablenotetext{b}{rms values for a 20 km s$^{-1}$ channel width for all lines except $^{13}$CO J=2-1 which uses 50 km s$^{-1}$ channel width.}
\tablenotetext{c}{Map obtained from \citet{1997ApJ...475L.107A}; observed with Owens Valley Radio Observatory.}
\end{deluxetable}

\begin{deluxetable}{ccccccc} 
\tablecolumns{7}
\tablewidth{0pt}
\tablecaption{Comparision of JCMT and Combined (Feathered) Maps\tablenotemark{a}}
\label{smafeather}
\tablehead{\colhead{} & \multicolumn{3}{c}{Total Flux (Jy km s$^{-1}$)\tablenotemark{b}} & \colhead{} &  \multicolumn{2}{c}{} \\ \cline{2-4}   \\\colhead{Transition} & \colhead{SMA} & \colhead{Feather} &\colhead{JCMT}& \colhead{}  & \colhead{Missing\tablenotemark{c}} &\colhead{Recovered\tablenotemark{d}}  \\ \colhead{($^{12}$CO)} & \colhead{} & \colhead{} &\colhead{} & \colhead{} &\colhead{($\%$)} &\colhead{($\%$)}  }
\startdata
J=2-1	&1870 $\pm$ 30		 &2380 $\pm$ 30 	&2320 $\pm$ 120 &&	20 $\pm$ 30&100 $\pm$ 30 \\
J=3-2	& 2820 $\pm$ 40		 &6350 $\pm$ 50  &8770 $\pm$ 120&&	70 $\pm$ 30& 70 $\pm$ 40\\
\enddata
\tablenotetext{a}{$^{13}$CO J=2-1 and $^{12}$CO J=1-0 single dish maps are not available}
\tablenotetext{b}{measurement uncertainty only; calibration uncertainty is 20$\%$ for SMA and the JCMT $^{12}$CO J=2-1 maps and 30$\%$ for the JCMT $^{12}$CO J=3-2 (Paper I). }
\tablenotetext{c}{Missing flux in the SMA only map relative to the JCMT map. Uncertainty calculation includes calibration uncertainty.}
\tablenotetext{d}{Recovered flux in the feathered map relative to the JCMT map. Uncertainty calculation includes calibration uncertainty.}
\end{deluxetable}

\begin{deluxetable}{ccccc} 
\tablecolumns{5}
\tablewidth{0pt}
\tablecaption{}
\label{dv}
\tablehead{\colhead{Region} & \colhead{Velocity FWHM\tablenotemark{a}}&  \multicolumn{2}{c}{Deconvolved Source Diameter\tablenotemark{b}} & \colhead{$M_{\rm{dyn}}$} \\ \colhead{} & \colhead{(km s$^{-1}$)}&  \colhead{($\arcsec$)} & \colhead{(pc)} & \colhead{(10$^{9}$ $M_{\odot}$)} }
\startdata
A (IC 694)& 325	&(1.7 $\pm$ 0.5) $\times$ (1.2 $\pm$ 1.0)	&380 $\times$ 270 & (3.4 $\pm$ 2.1)  	\\
A2 (IC 694 disk)\tablenotemark{c}&90 &-&-&- \\
B  (NGC 3690)&185&(2.0 $\pm$ 0.6) $\times$ (1.4 $\pm$ 1.0)	&450 $\times$ 310 & (1.3 $\pm$ 0.7)  	\\
C1 (Overlap)&80&(2.5 $\pm$ 0.8)$\times$ (1.6 $\pm$ 1.3)	&560 $\times$ 360 & (0.3 $\pm$ 0.2)  	\\
C2 (Overlap)&80&(4.3 $\pm$ 2.1) $\times$ (1.5 $\pm$ 1.4) &960 $\times$ 330 & (0.4 $\pm$ 0.3) 
\enddata
\tablenotetext{a}{Uncertainty in velocity FWHM is 10 km s$^{-1}$}
\tablenotetext{b}{Beam size is 2.2$\arcsec$ $\times$ 1.9$\arcsec$}
\tablenotetext{c}{Region A2 cannot be fit with a Gaussian because it is too close to region A}
\end{deluxetable}

\begin{deluxetable}{lcccc} 
\tablecolumns{5}
\tablewidth{0pt}
\tablecaption{CO Line Ratios of Arp 299\tablenotemark{a}}
\label{lineratios}
\tablehead{\colhead{Region} & \colhead{$^{12}$CO $\frac{J=3-2}{J=2-1}$\tablenotemark{b}} &  \colhead{$^{12}$CO $\frac{J=3-2}{J=2-1}$\tablenotemark{c}} &\colhead{$^{12}$CO $\frac{J=2-1}{J=1-0}$} & \colhead{$\frac{\mathrm{^{12}CO}}{\mathrm{^{13}CO}}$ J=2-1} }
\startdata
A 	&0.74 $\pm$ 0.21	&0.72 $\pm$ 0.20	&1.37 $\pm$ 0.39	&30.9 $\pm$ 4.4		\\
A2	&0.76 $\pm$ 0.22 	&0.58 $\pm$ 0.16	&1.32 $\pm$ 0.38	&12.6 $\pm$ 1.8		\\
B 	&0.73 $\pm$ 0.21	&0.73 $\pm$ 0.21	&1.66 $\pm$ 0.47	&27.3 $\pm$ 3.9		\\
C1 	&0.52 $\pm$ 0.15	&0.51 $\pm$ 0.15	&1.24 $\pm$ 0.35	&17.2 $\pm$ 2.4		\\
C2 	&0.65 $\pm$ 0.19	&0.60 $\pm$ 0.17	&1.23 $\pm$ 0.35	&11.6 $\pm$ 1.6	\\
\enddata
\tablenotetext{a}{All ratios are in terms of integrated brightness temperatures at the peak intensity value found in the $^{12}$CO J=3-2 map. Uncertainties include calibration errors only. All maps were degraded to a resolution of 3.6$\arcsec$ $\times$ 2.4$\arcsec$.}
\tablenotetext{b}{Using short spacing corrected maps}
\tablenotetext{c}{Using SMA only maps}
\end{deluxetable}

\begin{deluxetable}{cccc} 
\tablecolumns{7}
\tablewidth{0pt}
\tablecaption{Radex Results}
\label{radexsol}
\tablehead{\colhead{Region} & \colhead{$N$($^{12}$CO)} & \colhead{$T_{\rm{kin}}$} & \colhead{$n$(H$_{\rm{2}}$)}  \\ \colhead{} & \colhead{(cm$^{-2}$)} & \colhead{(K)} & \colhead{(cm$^{-3}$)} }
\startdata
A 	&1-8 $\times$ 10$^{18}$	&10-500	&$>$10$^{2.5}$\\
A2  	& 2-30 $\times$ 10$^{18}$ 	&20-500	&$>$10$^{2.5}$\\
B 	&1-6 $\times$ 10$^{18}$		&10-1000	&$>$10$^{2.5}$\\
C1 	&1-4 $\times$ 10$^{18}$	&10-50	&10$^{3}$-10$^{4.5}$\\
C2 	&2-9 $\times$ 10$^{18}$	&10-200	&10$^{3}$-10$^{5}$\\
\enddata
\end{deluxetable}

\begin{deluxetable}{lccccc} 
\tablecolumns{6}
\tablewidth{0pt}
\tablecaption{Gas Mass and Depletion Times}
\label{gasmass}
\tablehead{\colhead{Region} & \colhead{$\dot{M}_{\rm{SFR}}$} & \colhead{$^{12}$CO J=2-1 Flux\tablenotemark{a}} & \colhead{$L$$_{\rm{CO (2-1)}}$\tablenotemark{bc}} & \colhead{$M$(H$_{\rm{2}}$)\tablenotemark{c}} & \colhead{t$_{\rm{depl}}$\tablenotemark{c}} \\ \colhead{} & \colhead{($M_{\odot}$ yr$^{-1}$)} & \colhead{(Jy km s$^{-1}$)} &\colhead{(10$^{9}$ K km s$^{-1}$ pc$^{2}$)}&  \colhead{(10$^{8}$ $M_{\odot}$)} & \colhead{(Myr)}}
\startdata
A 	& 35.8	&920 $\pm$ 100	& 1.2 $\pm$ 0.2 	&7 $\pm$ 1	& 27 $\pm$ 5	\\
A2	&  $<$7.2\tablenotemark{d}	& 320 $\pm$ 20& 0.4 $\pm$ 0.1 & 2.3 $\pm$ 0.5 & $>$43 \\
B	& 19.3	& 390 $\pm$ 40	& 0.5 $\pm$ 0.1	& 3.0 $\pm$ 0.6	&21 $\pm$ 4	\\
C = C1 + C2	& 9.3	& 510 $\pm$ 30	& 0.7 $\pm$ 0.1	& 4.0 $\pm$ 0.7	& 60 $\pm$ 10 \\
Total		&71.5 &2380 $\pm$ 20	&3.1 $\pm$ 0.6  	& 18 $\pm$ 4	&34 $\pm$ 7\\
\enddata
\tablenotetext{a}{measurement uncertainty only; calibration uncertainty is 20$\%$}
\tablenotetext{b}{Using $D_{\rm{L}}$ = 46 Mpc}
\tablenotetext{c}{Uncertainty is calculated using the calibration uncertainty}
\tablenotetext{d}{$\dot{M}_{\rm{SFR}}$ calculated assuming the region contains 10$\%$ the total infrared luminosity of Arp 299.}
\end{deluxetable}

\bibliography{references2}
\bibliographystyle{apj}

\end{document}